\documentclass[a4paper,fleqn,usenatbib]{mnras}

\usepackage{newtxtext,newtxmath}

\usepackage[T1]{fontenc}
\usepackage{ae,aecompl}

\usepackage{hyperref}
\usepackage{graphicx}	
\usepackage{amsmath}	
\usepackage{amssymb}	
\usepackage{epstopdf}
\usepackage{booktabs}
\usepackage{float}
\usepackage{subcaption}
\usepackage{multicol}
\usepackage{rotating}

\title[Distances to Three SNRs]{Distances to Supernova Remnants G$20.4+0.1$, G$24.7-0.6$ and G$28.6-0.1$ and New Molecular Cloud Associations}

\author[S. Ranasinghe]{
S. Ranasinghe\thanks{E-mail: syranasi@ucalgary.ca}
and D. A. Leahy\\
Department of Physics \& Astronomy, University of Calgary, Calgary, Alberta T2N 1N4, Canada
}

\date{Accepted XXX. Received YYY; in original form ZZZ}

\pubyear{2017}

\begin{document}
\label{firstpage}
\pagerange{\pageref{firstpage}--\pageref{lastpage}}
\maketitle

\begin{abstract}
Accurate distances to supernova remnants (SNRs) are crucial in determining their size, age, luminosity and evolutionary state. 
To determine distances, we chose three SNRs from the VLA Galactic Plane Survey (VGPS) for extraction of HI absorption spectra. 
Analysing HI absorption spectra, $^{13}$CO emission spectra, and HI and $^{13}$CO channel maps, kinematic velocities (or their limits) to the three SNRs were calculated. The three SNRs are probably associated with molecular clouds and the new distance to G$20.4+0.1$, G$24.7-0.6$ and G$28.6-0.1$ are $7.8 \pm0.5$ kpc, $3.8\pm 0.2 $ kpc and $ 9.6 \pm 0.3 $ kpc, respectively.
\end{abstract}

\begin{keywords}
ISM: supernova remnants -- radio continuum: ISM -- radio lines: ISM
\end{keywords}



\section{Introduction} \label{sec:intro}

\indent For the study of a supernova remnant (SNR), distance as a basic physical parameter is crucial in order to determine its size, luminosity and age. 
SNRs are an important source of energy and element-enriched ejecta input for the interstellar medium (ISM). 
They are associated with several highly active phenomena including Anomalous X-ray Pulsars (AXP), Soft Gamma-ray Repeaters (SGR), Pulsar Wind Nebulae (PWNe) and non-thermal X-ray and very high energy emission. 
Distance determination of a SNR leads to association with other objects and may help constrain the mass range of the progenitor star and type of supernova resulting in the remnant. \\
\indent However, determination of the distances to SNRs has been a difficult task. 
A common method of obtaining distances is by analysing HI absorption spectra along with associated HII regions and molecular clouds. 
Due to observational constraints, constructing reliable HI absorption spectra has been difficult in the past. However, with recent improvements in observations with better sensitivity and higher resolution, the determination of accurate distances to SNRs have become possible. \\
\indent To construct HI absorption spectra, we use the method developed by \cite{Leahy2010}. We obtain the distance to SNRs by analysing HI absorption and ${^13}$CO spectra with the HI and ${^13}$CO channel maps \citep{2017Ranasinghe}. \\
\indent In Section \ref{sec:DA}, we discuss the data and software used for this analysis, construction of HI absorption spectra, determination of kinematic distances and error analysis.
The results for the three selected SNRs are presented in Section \ref{sec:results} and in Section \ref{sec:Disconc} a summary is given. 

\section{Data Analysis} \label{sec:DA}

\subsection{Data and Software}

\indent  HI emission-line data and 1420 MHz continuum data used here are from the VLA (Very Large Array) Galactic Plane Survey (VGPS) \citep{Stil}. 
The source of $^{13}$CO line data is the Galactic Ring Survey taken with the 
14 m telescope of the Five College Radio Astronomical Observatory (FCRAO)  \citep{Jackson}. 
We construct source and background spectra using MEANLEV, a software program in the DRAO EXPORT package. This program allows one to extract on and off spectra defined by spatial boundaries and by user-specified threshold T$_{B}$ levels.

\subsection{HI Absorption and $^{13}$CO Emission Spectra }

\indent HI absorption spectra were made using the method of \cite{Leahy2010}. 
The equation of radiative transfer is $\frac{dI_{\nu}}{ds} = j_{\nu} - \kappa_{\nu}I_{\nu}$, with  specific intensity I$_{\nu}$, element of distance 
along the LOS (line-of-sight) $ds$ , emission coefficient $j_{\nu}$ and linear absorption coefficient $\kappa_{\nu}$. 
From the radiative transfer equation, it follows the HI absorption spectrum is given by:
\begin{equation} \label{eq1}
e^{-\tau_{\textrm{v}}} -1 = \frac{T_{B,\textrm{on}}(\textrm{v})-T_{B,\textrm{off}}(\textrm{v})}{T_{B,\textrm{on}}^{C}-T_{B,\textrm{off}}^{C}},
\end{equation}
where $T_{B}$ is the brightness temperature, `on' refers to the LOS to the source and `off' refers to the LOS to the background. A derivation of equation \ref{eq1} is given by \cite{Leahy2010}.
The optical depth vs. frequency (or equivalently radial velocity, v) 
$\tau_{\textrm{v}}$, is the HI absorption spectrum and is plotted here as $e^{-\tau_{\textrm{v}}}$. \\
\indent The brightest continuum regions for each SNR were chosen for the `on' position. The 'off' regions were chosen using the HI channel maps to avoid
bright HI features which could result in false HI absorption features. 
Using the MEANLEV program (from the DRAO export package) allows the source and background to be close to each other or adjacent. \\
\indent We used \textbf{the same source and background regions as for the HI spectra} to extract $^{13}$CO emission spectra. 
The $^{13}$CO channel maps were examined to determine whether a molecular cloud could be associated with the SNR.  
The border of the SNR was taken as the edge of the continuum radio emission. 
Then the reality of association with the SNR was based on the spatial correlation between the SNR border and the molecular cloud border.

\subsection{Distance Determination}
\indent Once a radial velocity is identified with an object, the distance to that object can be determined. We use a circular rotation model, $V(R)$, for the Galaxy, so that radial velocity, $V_{r}$ is: 
\begin{equation} \label{eq2}
 V_{r} =  R_{0}\sin(l) \left(\frac{V(R)}{R}- \frac{V_{0}}{R_{0}}\right),
\end{equation}
where $l$ is the galactic longitude and $V(R)$ is the orbital velocity at the Galacto-centric distance R. 
The rotation curve provides $V(R)$ for any $R$, hence $V(R)/R$, 
and thus $R$ can be found.
The distance, $d$, from the Sun for an object at R with Galactic longitude, $l$, 
satisfies
$R^{2} = R_{0}^{2}  + d^{2}- 2R_{0}d\cos(l)$. This yields
\begin{equation} \label{eq3}
d    =  	R_{0}\cos(l) \pm \sqrt{R^2  - R_{0}^2 \sin^{2}(l)}.
\end{equation}
The existence of two possible distances for locations inside the solar circle, corresponding 
to the plus and minus signs above, is called the Kinematic Distance Ambiguity (KDA).

For each SNR, there may be a more than one overlapping or adjacent molecular clouds.
To verify an particular cloud is associated with the SNR one can 
use definitive indicators of shocked molecular gas: masers or broad CO lines.
However for the 4 SNRs discussed here, there are no maser or molecular cloud associations previously known.\\
\indent To determine distance to a SNR from its radial velocity, the appropriate rotation curve for that region of the Galaxy is needed. 
We use $V(R)$ from \cite{Reid2014} which is derived for the region we are studying. 
This uses the formula of \cite{Persic} (called the Universal Rotation Curve or URC). 
The distance of the Sun to the Galactic center is $R_{0} = 8.34 \pm 0.16$ kpc; 
the orbital velocity of the sun is $V_{0} = 241 \pm 8$ km s$ ^{-1}$; 
the radius encompassing $83\%$ of the light is $R_{opt} = R_{0} \cdot (0.90 \pm 0.006)$; the circular velocity at $R_{opt}$ is $V(R_{opt}) = 241  \pm  8$;
the velocity core radius $a = 1.5$ and  the disk fraction is $\beta = 0.72$ \citep{Persic}.

\subsection{Error Analysis}

\indent Once HI absorption spectra for the SNRs were constructed, we investigated the individual HI channel maps to verify whether each absorption feature was real or an artifact. 
We based our final conclusions by comparing the HI channel maps and the HI spectra. \\
\indent In some cases, we see calculated values of e$^{-\tau} < 0$ or e$^{-\tau} > 1$ (e.g. the blue line in the lower half of the top two panels of Figure \ref{fig:5}). 
These features are due to HI emission which is in one of, but not both, the background region or source region. 
For example, Figure \ref{fig:6A} bottom panel (region 2) shows a case where there is excess HI emission, i.e. excess
$T_{B,\textrm{off}}$, in the background region (the part of
the red box outside the SNR). 
This shows up in the HI spectrum (see equation \ref{eq1}) as a negative contribution to 
e$^{-\tau}$, i.e. a false absorption feature.  
Conversely, there may be excess 
HI emission (excess $T_{B,\textrm{on}}$) in the source region. 
This excess HI emission in the source region results in a positive contribution to 
e$^{-\tau}$ (equation \ref{eq1}), and, if strong enough, yields a value e$^{-\tau} > 1$. 
Excess HI emission in the the background region,  
if strong enough, can yield a value e$^{-\tau} < 0$.
Disregarding a feature in an HI spectrum was done after thoroughly investigating the HI channel maps.\\
\indent	The error estimates for the distances  were done as follows. 
Because the equations to determine distance from radial velocities are non-linear, we calculated a set of distances for our best fit parameters and
for upper and lower limits of the parameters. 
The three parameters are R$_{0}$, V$_{0}$ and observed radial velocity V$_{r}$, resulting in a set of $3^{3}=27$ parameters and distances. 
Here the errors for R$_{0}$ and V$_{0}$ were taken from \cite{Reid2014} ($\pm0.16$ kpc for R$_{0}$ and $\pm8$ km s$^{-1}$ for V$_{0}$).
The measurement error in  V$_{r}$ was taken as $\pm 2.4$ km s$^{-1}$ based on the channel maps. 
As an estimate of the peculiar gas motion we take 4.7 km s$^{-1}$ based on the difference of measured tangent point velocities and the URC. 
Added in quadrature the net error in V$_{r}$ is 5.3 km s$^{-1}$. 
The standard deviation of the calculated distances (27 values per SNR) yields our estimate of the error in the distance. For the tangent point distance error, we use the corresponding $\pm5.3$ km s$^{-1}$ of the tangent point velocity. However, if there is an association of a molecular cloud at the tangent point, we take the FWHM of the $^{13}$CO as the error. 
	
\section{Results} \label{sec:results}

\begin{figure}
\includegraphics[scale=.34]{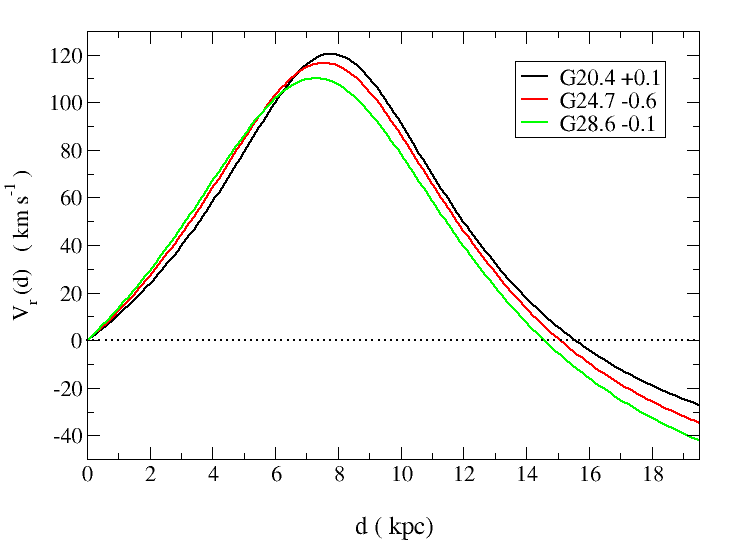}\par 
\caption{The radial velocity V$_{r}$ as a function of the distance (d) at each longitude corresponding to the three SNRs.}
\label{fig:1A}
\end{figure}  

\begin{figure}
\centering
\includegraphics[scale=.45]{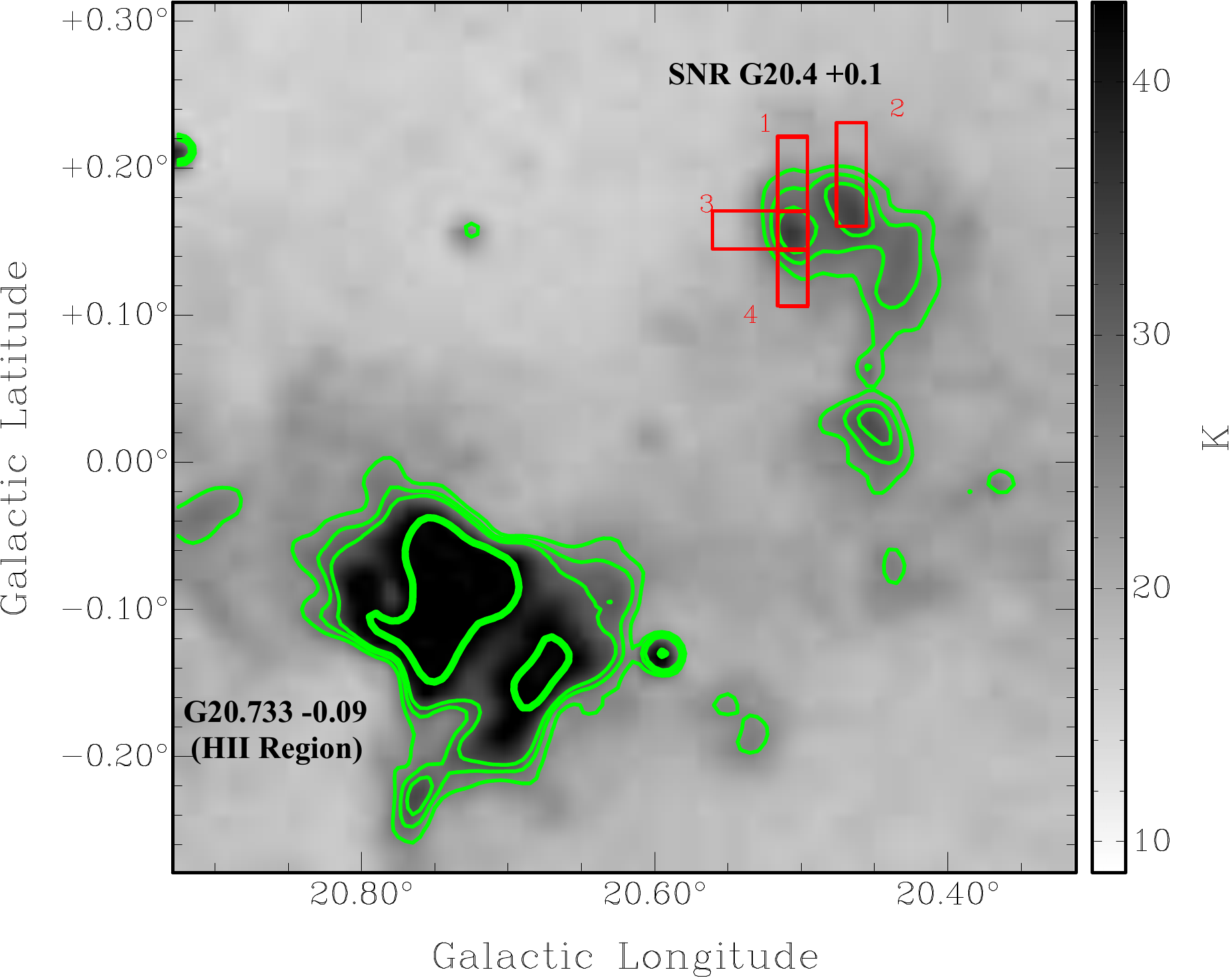}
\caption {SNR G$20.4+0.1$ 1420 MHz image with green contours at 25, 28, 31 and 50 K. The red boxes are the areas used to extract HI and $^{13}$CO source and background spectra.}
\label{fig:1}
\end{figure}

\begin{figure}
\includegraphics[scale=.33]{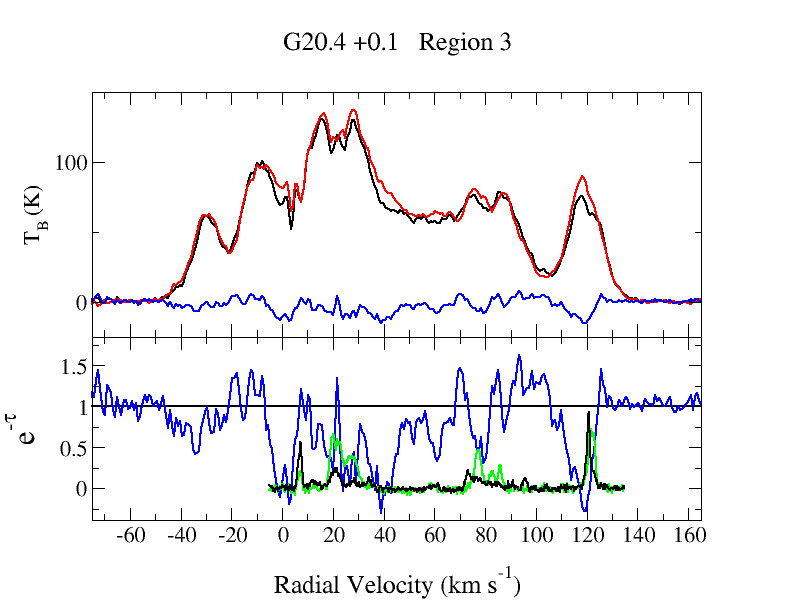}\par
\caption{G$20.4+0.1$ region 3 spectrum.  Top panel shows the HI emission spectra where the source spectrum (black), background spectrum (red) and difference (blue).
The bottom panel shows the HI absorption spectrum in blue, the $^{13}$CO source spectrum in green and the $^{13}$CO background spectrum in black.}
\label{fig:2}
\end{figure}  

\begin{figure}
  \includegraphics[scale= 0.45]{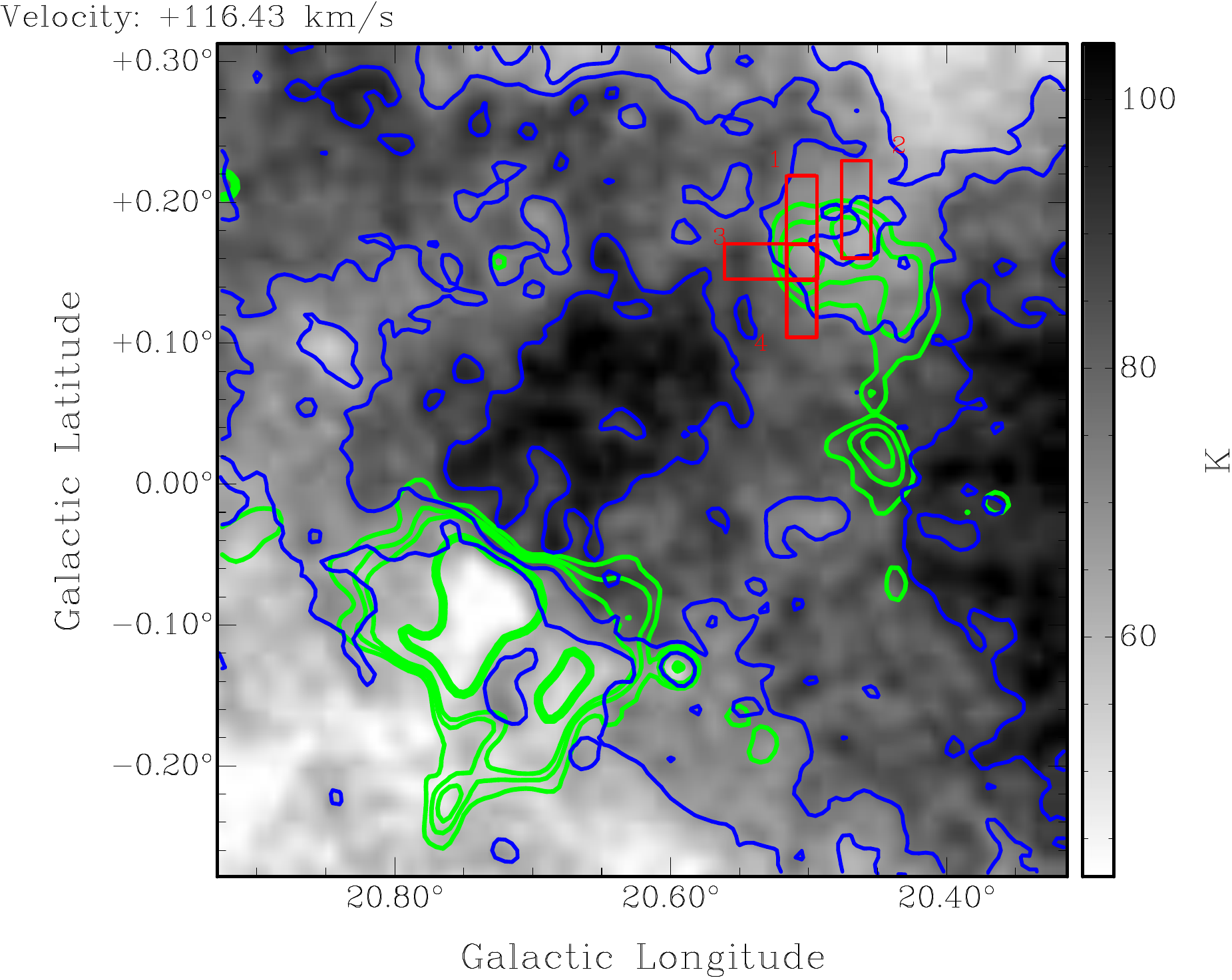}
\caption{G$20.4+0.1$ HI channel map +116.43 $ \text{km s}^{-1}$. The blue HI contours at 64 and 75 K and the green continuum contours at 25, 28, 31 and 50 K.}
\label{fig:3A}
\end{figure}

\begin{figure}
  \begin{subfigure}{\linewidth}
  \includegraphics[scale= 0.45]{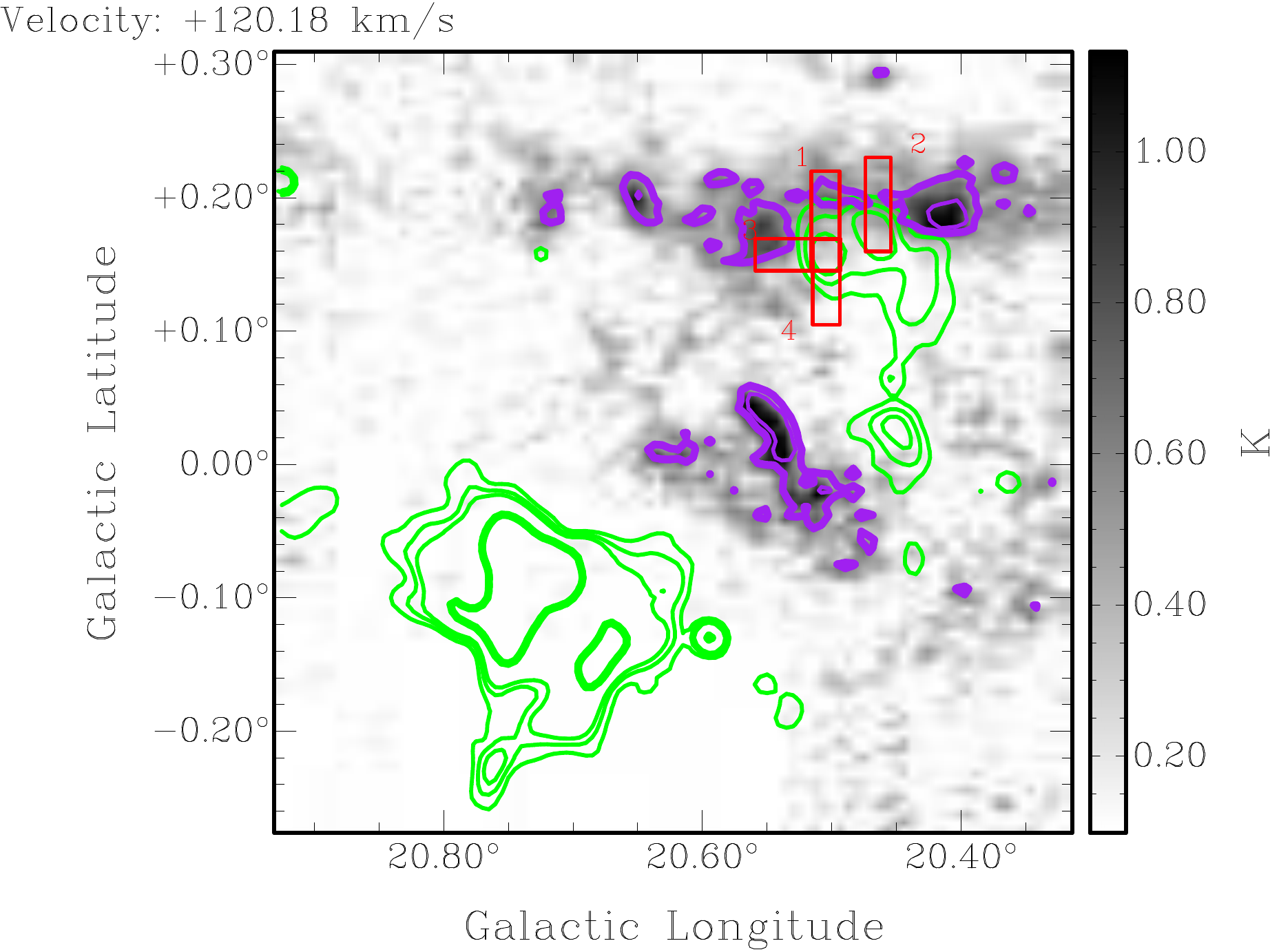}\hfill
  \includegraphics[scale= 0.45]{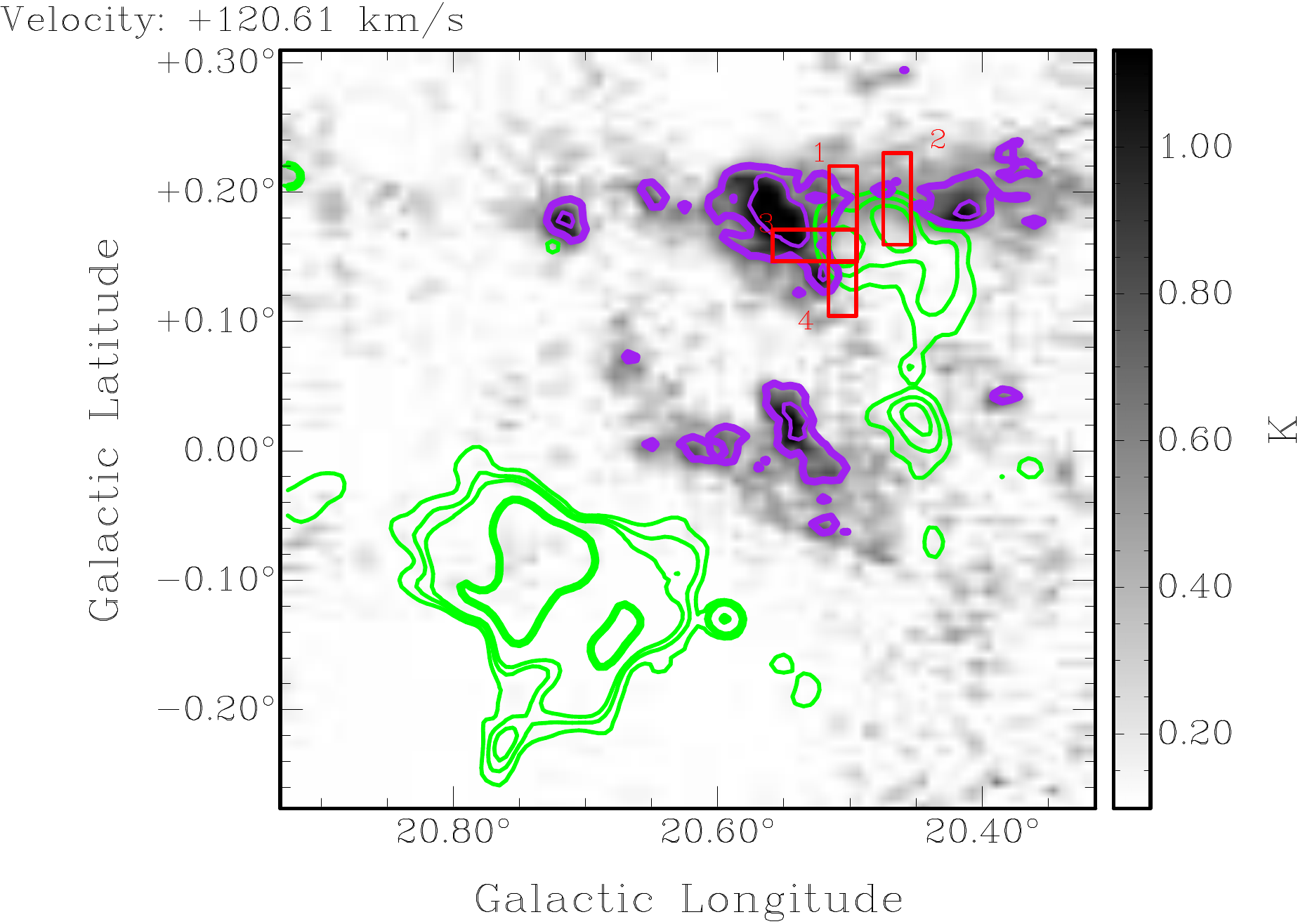}
  \end{subfigure}\par\medskip
  \begin{subfigure}{\linewidth}
  \includegraphics[scale= 0.45]{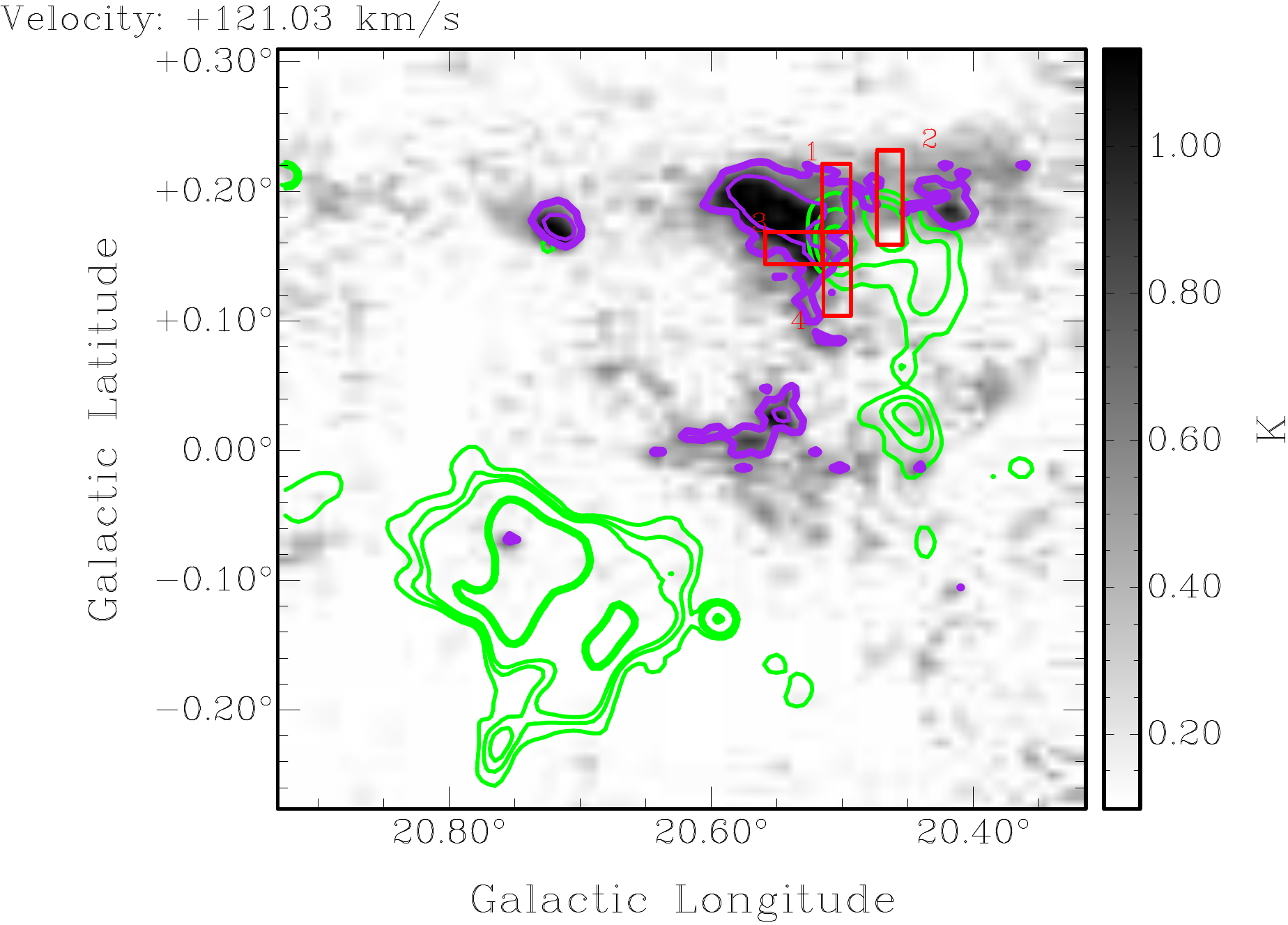}
  \end{subfigure} 
\caption{G$20.4+0.1$ $^{13}$CO channel maps +120.6 to +121.03  $ \text{km s}^{-1}$. The purple $^{13}$CO contours at 0.6 and 1 K. Thegreen continuum contours at 25, 28, 31 and 50 K.}
\label{fig:3}
\end{figure}

\subsection{G$20.4+0.1$} \label{G20}

\indent Also known as G$20.47 +0.16$, the SNR was first discovered by \cite{BroganGelfand2006} using the 90 cm multi-configuration VLA survey of the Galactic plane. 
The SNR has a spectral index that is $\sim$ 0.1 with a high degree of uncertainty.
The $7^\prime \times 7^\prime $ SNR is in a complex region with possible overlapping HII regions and radio sources. 
Figure \ref{fig:1} shows 1420 MHz image of G$20.47 +0.16$.
For this SNR there is no known distance, but the HII region G$20.733 -0.09$, to the 
lower-left of the SNR, is placed at a distance of 11.9 kpc \citep{Quireza2006}.\\
\indent  Extracting HI absorption spectra for the SNR was difficult due to random HI clouds confusing the separation of the source and background. 
The best results were obtained using the brightest region as the source with three different background regions (see regions 1, 3 and 4 in Figure \ref{fig:1}).
We investigated the HI channel maps to verify or falsify features in the absorption spectra.\\
\indent Inspecting the HI spectra, it is seen that they are dominated by noise. Hence we present the spectrum with the least amount of noise. From Figure \ref{fig:2} it is seen there may be absorption present up to the tangent point. The maximum tangent point velocity for the SNR is at 120.4 km s$^{-1}$ (Figure \ref{fig:1A})
This conclusion is strengthened by examining the HI channel maps (Figure \ref{fig:3A}).  
At a radial velocity of $\sim$116 km s$^{-1}$, the absorption feature matches the shape of the continuum intensity. 
Hence the SNR is at least at a distance of the tangent point, 7.8 kpc. \\
\indent Figure \ref{fig:3} shows the $^{13}$CO channels near the tangent point velocity ($\sim120$ $ \text{km s}^{-1}$). 
It is seen that there is a molecular cloud with boundary that closely matches the 
northern border of the SNR. The conclusion is that there is a 
molecular cloud at the tangent point probably associated with the SNR.
This observation constrains the distance to be very nearly the tangent point distance, 7.8 kpc.\\
         
\subsection{G$24.7-0.6$} \label{G24}

\begin{figure}
\centering
\includegraphics[scale=.42]{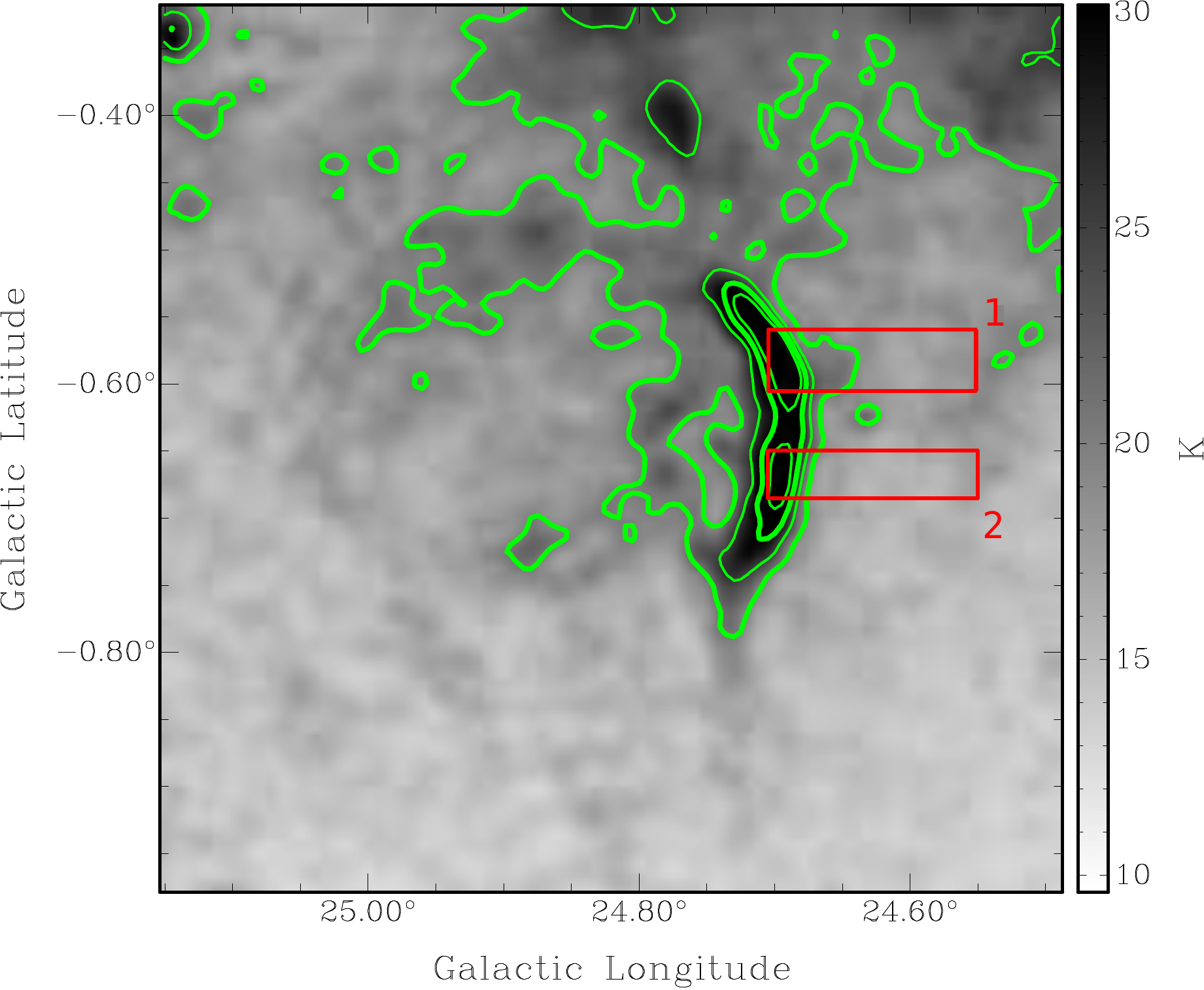}
\caption {SNR G$24.7-0.6$ 1420 MHz image with green contours at 20, 25, 30 and 35 K.
 }
\label{fig:4}
\end{figure}

\begin{figure}
\includegraphics[scale=.33]{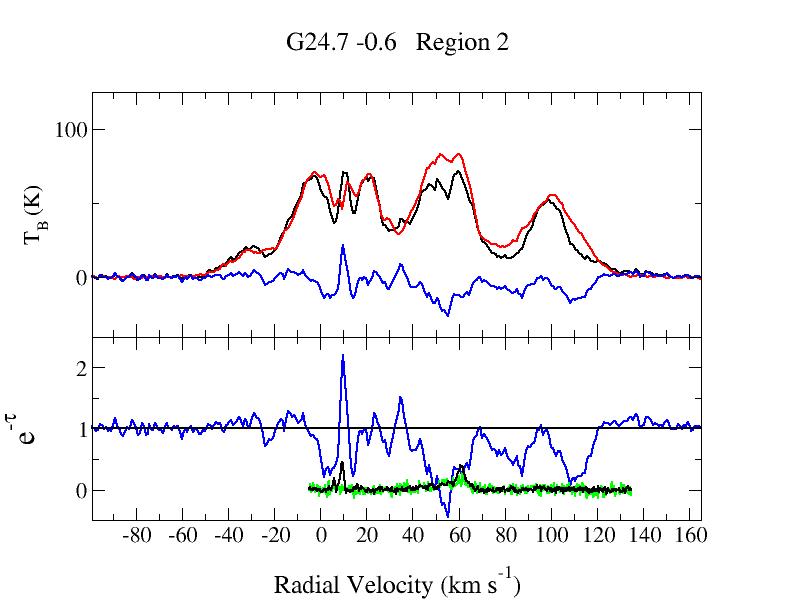}
\caption{G$24.7-0.6$  Region 2 spectrum.  Top panel shows the HI emission spectra where the source spectrum (black), background spectrum (red) and difference (blue).
The bottom panel shows the HI absorption spectrum in blue, the $^{13}$CO source spectrum in green and the $^{13}$CO background spectrum in black.
}
\label{fig:5}
\end{figure} 

\begin{figure}
\centering
\includegraphics[scale=.44]{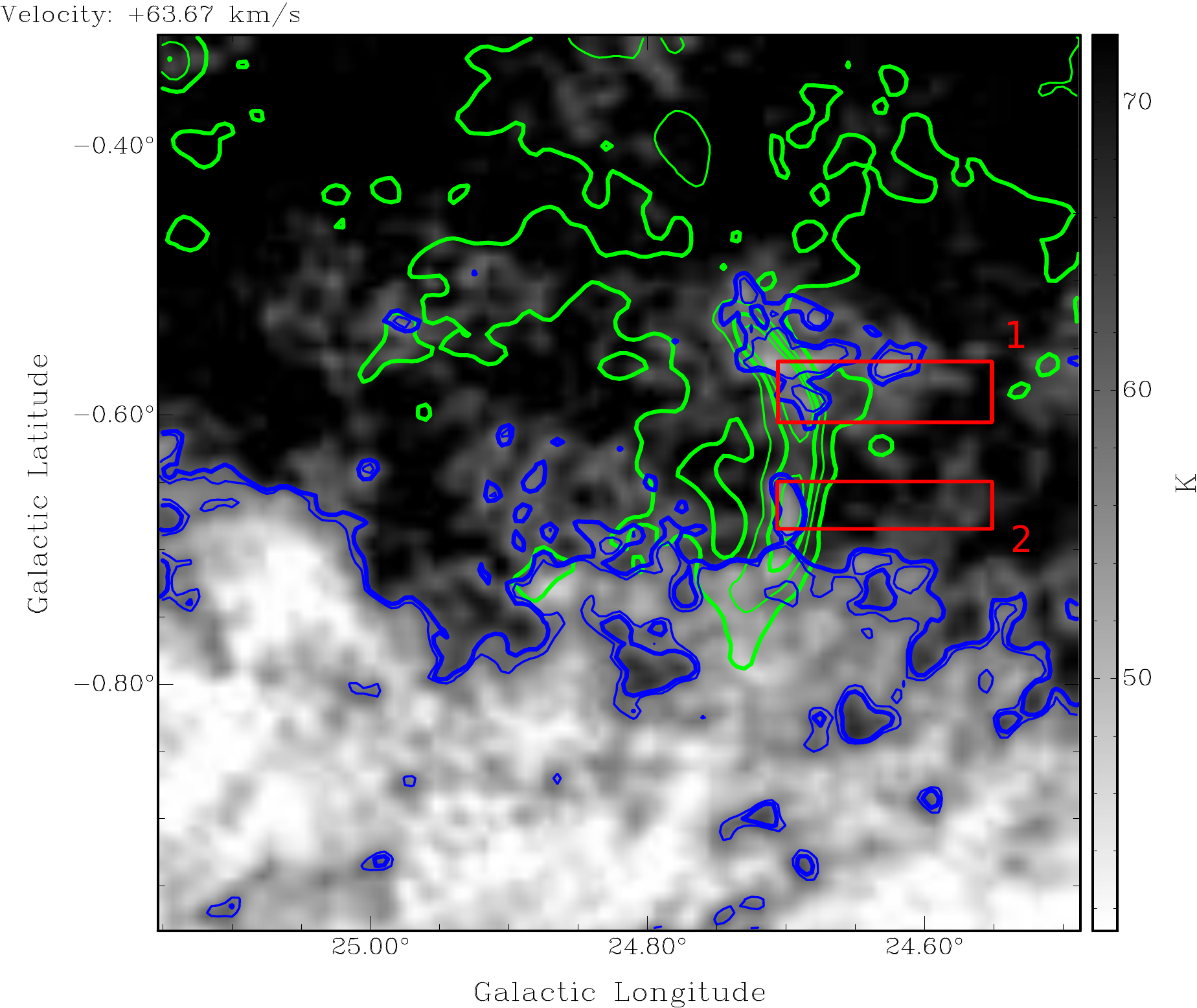}
\includegraphics[scale=.45]{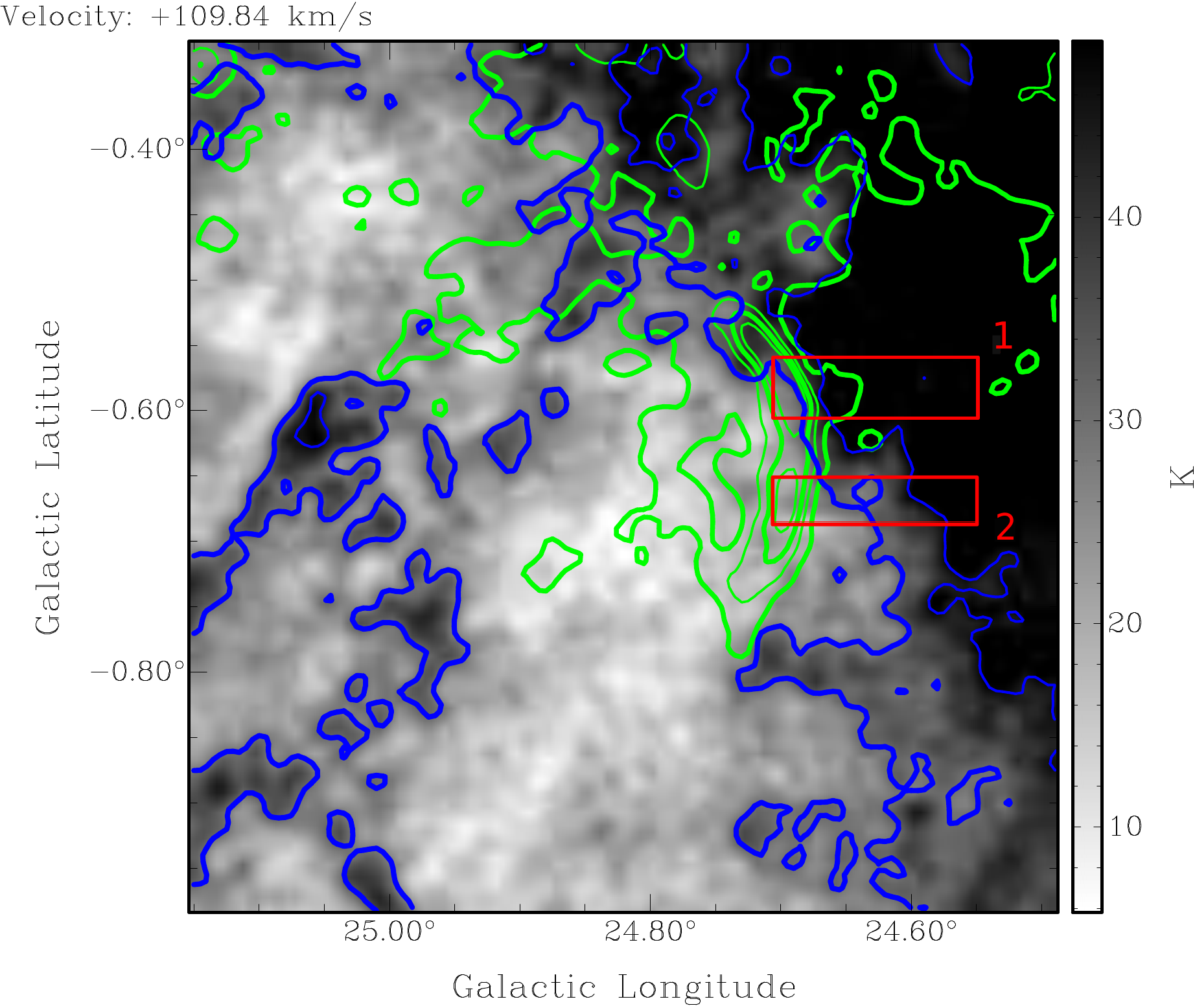}
\caption {G$24.7-0.6$ HI channel maps  +63.67 and 109.84$ \text{km s}^{-1}$. The blue HI contours at 58 and 60 K for channel map +63.67$ \text{km s}^{-1}$ \& 30 and 45 K for channel map 109.84$ \text{km s}^{-1}$. The green continuum contours at 20, 25, 30 and 35 K.}
\label{fig:6A}
\end{figure}

\begin{figure}
  \includegraphics[scale= 0.44]{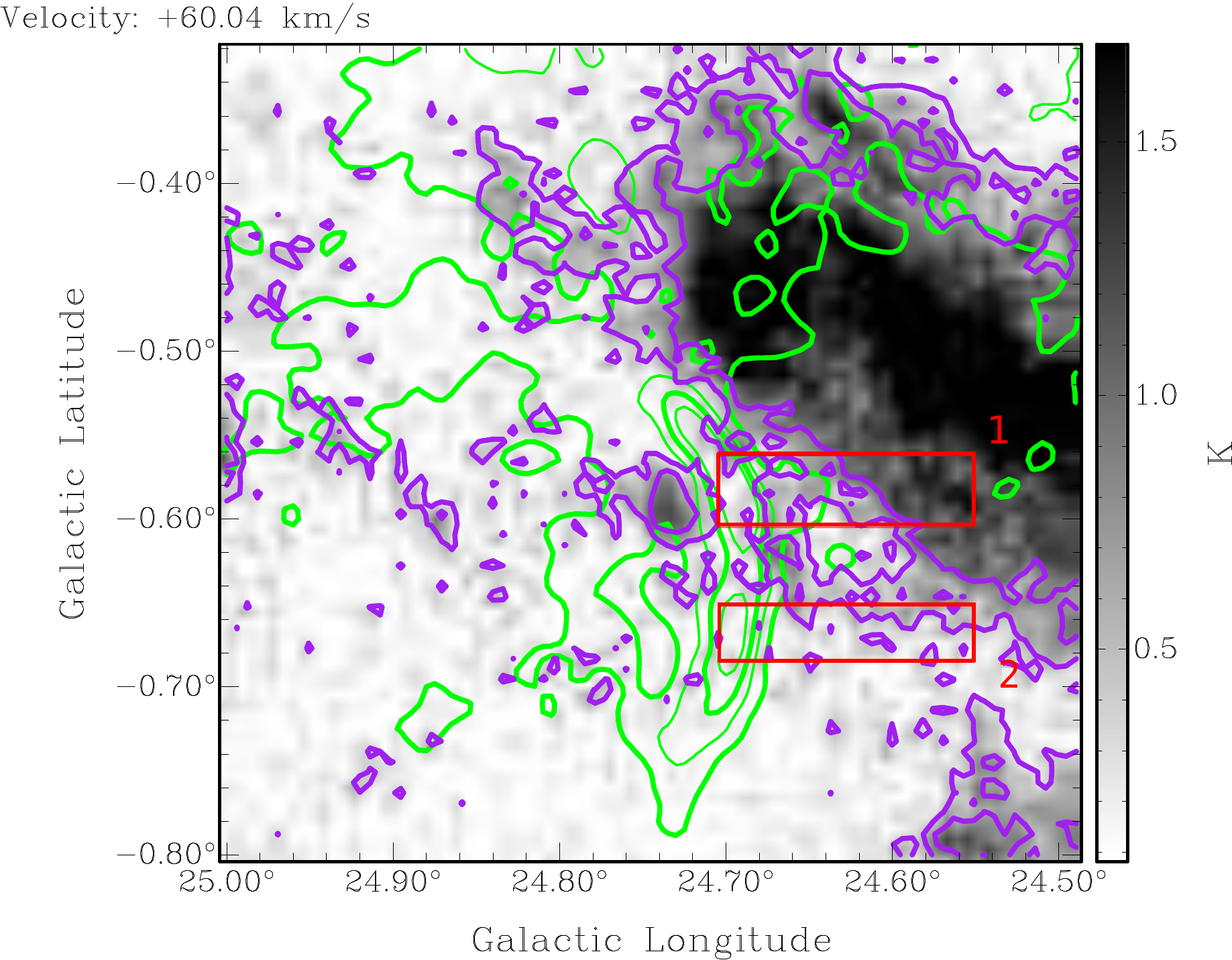}\hfill
  \includegraphics[scale= 0.44]{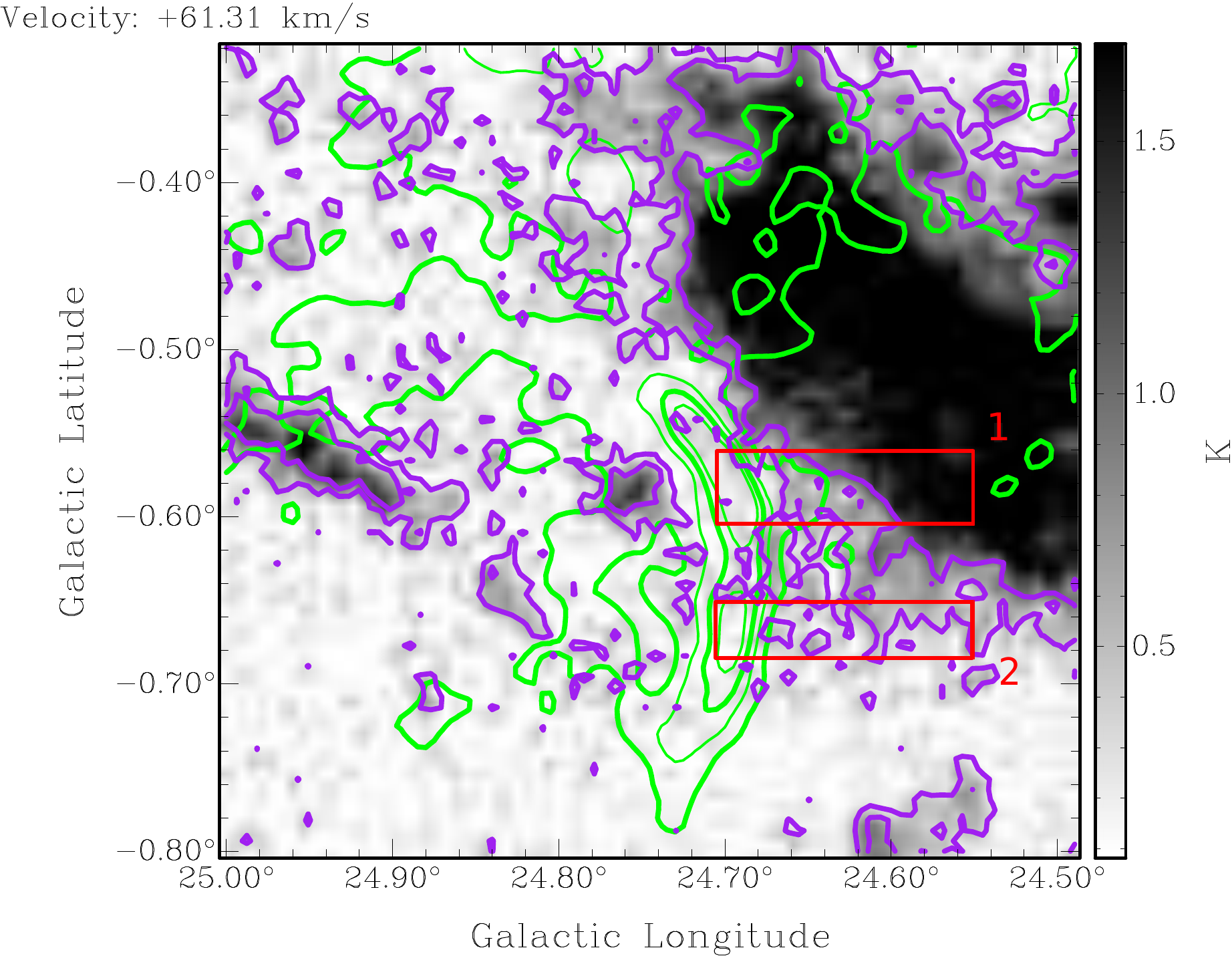}
\caption{G$24.7-0.6$ $^{13}$CO channel maps +60.04 and +61.31 $ \text{km s}^{-1}$. The purple $^{13}$CO contours are at 0.4 and 0.8 K. The green continuum contours at 20, 25, 30 and 35 K.}
\label{fig:6}
\end{figure}

\begin{figure*}
  \includegraphics[scale= 0.55]{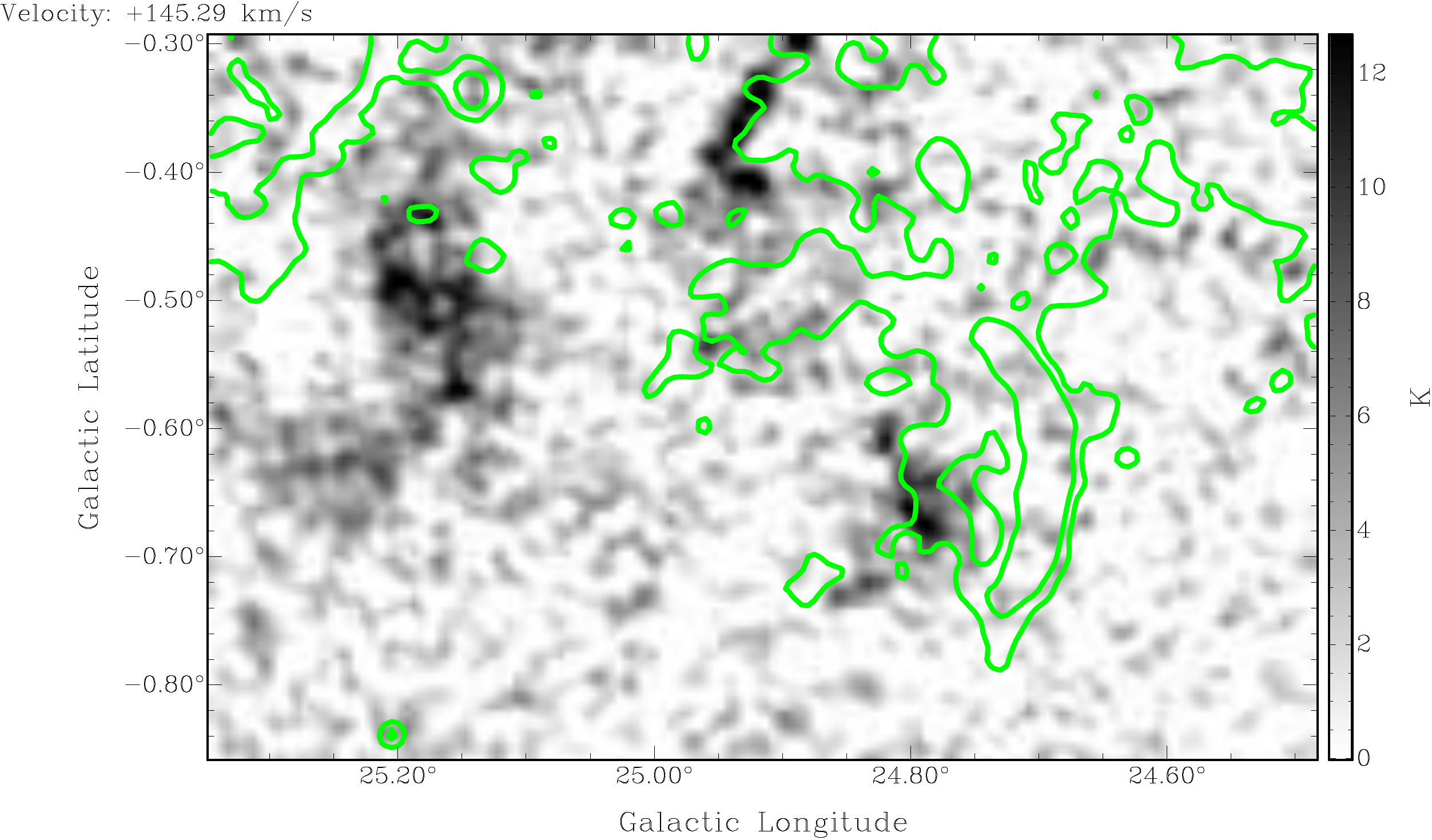}
\caption{G$24.7-0.6$ HI  +145.29 $ \text{km s}^{-1}$ channel map. The green continuum contours at 20 and 25 K.}
\label{fig:6C}
\end{figure*}

\indent The SNR was first observed by \citet{Clark1975} at 408 MHz and 5000 MHz. 
They reported the partial shell appearing at both frequencies and a spectral index of 0.49. 
Based on the 1465 MHz data of G$24.7 -0.6$, the estimated diameter is $\sim18^{\prime}$ for the incomplete shell \citep{Dubner1993}. 
\cite{KooHeiles1991} discussed the possibility of a velocity HI gas between 143 and 176 $ \text{km s}^{-1}$  implying an expanding HI shell. 
However, since the emitting region is much larger than the SNR, the emission is more likely to be associated with the nearby HII region complex. 
They estimated the distance to the SNR to be 9.3 kpc by assuming the SNR to
be associated with a nearby HII region.

Figure \ref{fig:4} shows 1420 MHz image of G$24.7 -0.6$. 
Regions 1 and 2 were chosen for extraction of HI and CO spectra.  
The absorption profiles for the two regions are similar, 
hence, we present the spectrum of region 2 in Figure \ref{fig:5}. 
This was used as a reference for the investigation of the HI channel maps. 

The HI spectra indicate absorption up to the tangent point. 
The individual HI channel maps do not confirm this.
The channel maps near 110 km s${^{-1}}$ show HI just outside the eastern rim of the SNR 
(Figure \ref{fig:6A} bottom panel) which matches the shape of the rim. 
This is clearly HI emission outside the SNR rather than HI absorption because the shape of the depression in HI  point does not correlate with the shape of the continuum intensity. 
None of the other channels between $\sim 100$ and $\sim120$ $\text{km s}^{-1}$
show clear evidence of absorption. 
Thus we conclude there is not absorption at the tangent point velocity. 
Since the SNR is a relatively weak continuum source, the lack of clear absorption at the
tangent point does not prove the SNR is located at the near distance.
However, there is strong HI emission between 90 and 110 km s${^{-1}}$ (top panel of Figure \ref{fig:5}). 
For most other SNRs that we have examined the strong HI emission is associated with absorption if the SNR is beyond the tangent point. E.g., SNRs \citep{2017Ranasinghe} such as G 31.9+0.0, brighter than G24.7-0.6,
and G54.4-0.3, fainter than G24.7-0.6, both show absorption.
Thus lack of absorption in presence of strong HI emission is a strong indicator (but not proof)
that the SNR is this side of the tangent point.

 Absorption features are seen in the HI absorption spectrum between velocities of 48 and 65 $\text{km s}^{-1}$.  
Figure \ref{fig:6A} (top panel) shows the HI channel map for 64 km s$^{-1}$. In this case, the depression in HI emission coincides
with the brightest continuum emission, verifying that the depression is caused by absorption.
The highest velocity with confirmed real HI absorption is 64 km s$^{-1}$.
The lack of absorption at velocities greater than 64 km s$^{-1}$ means either that
the SNR is at the near kinematic distance, or that there is no cold HI along the 
line-of-sight between the near kinematic distance and far kinematic distance.

Figure \ref{fig:6} shows two $^{13}$CO channel maps at 60 to 62 km s$^{-1}$.
There is a  good match between the edge of the molecular cloud and the north-western border of the SNR.  
The SNR appears to be in contact with the molecular cloud. 
Thus there is a probable association of the SNR with the molecular cloud, yielding
a SNR kinematic velocity of $\sim$60  to  62 $ \text{km s}^{-1}$.
This places the SNR either at the near distance of 3.8 kpc or at the far distance of 11.3 kpc, corresponding to the velocity of V$_{r} =$ 60.67 $ \text{km s}^{-1}$. 
We searched for HI self-absorption in the region around the SNR, but did not find any.
HI self-absorption is expected if there is emitting HI at the far distance which is 
absorbed by HI at the near distance, thus could be used to confirm a far distance.  
We favour the near distance of 3.8 kpc, because of the lack of absorption for any velocity higher than 64 km s${^{-1}}$.

\indent \cite{KooHeiles1991} mentioned high velocity HI in the area. 
In the VGPS HI data, high velocity HI is clearly detected.
The maximum radial velocity where there is any high velocity HI is around 154 km s${^{-1}}$.  
We show the VGPS channel map in the region around G24.7-0.6 at a velocity of 145 km s${^{-1}}$
(Figure \ref{fig:6C}), where an HI cloud overlapping the SNR is seen clearly. 
In comparing the channel maps from 135 to 150 km s${^{-1}}$ we find the position of the HI cloud does not shift with velocity. 
However, the positions of the other clouds in this area of sky do not shift either. 
The high velocity cloud overlapping G24.7-0.6 could be moving radially outward from the SNR, but this would not explain the channel images of the other clouds in the region.
If the different high velocity HI clouds are related, the flow direction is consistent 
being radially away from the observer over the whole region
rather than a flow radially directed away from the center of G24.7-0.6.
The concentrations of the high velocity HI gas, other than the cloud centred on G24.7-0.6, appear to be roughly consistent with the positions of the HII regions. 
Our conclusion is that a high velocity HI cloud is possibly associated with the SNR. 
However, the spatial distribution of the larger set of clouds in the region makes it 
also possible that the HI is associated with the complex of HII regions. 
Another possibility is that the high velocities may have some other origin.
In our opinion, the current data makes it difficult to decide among these options for
the origin of the high velocity HI.\\

\subsection{G$28.6-0.1$}  \label{G28}

\begin{figure}
\centering 
\includegraphics[scale=.42]{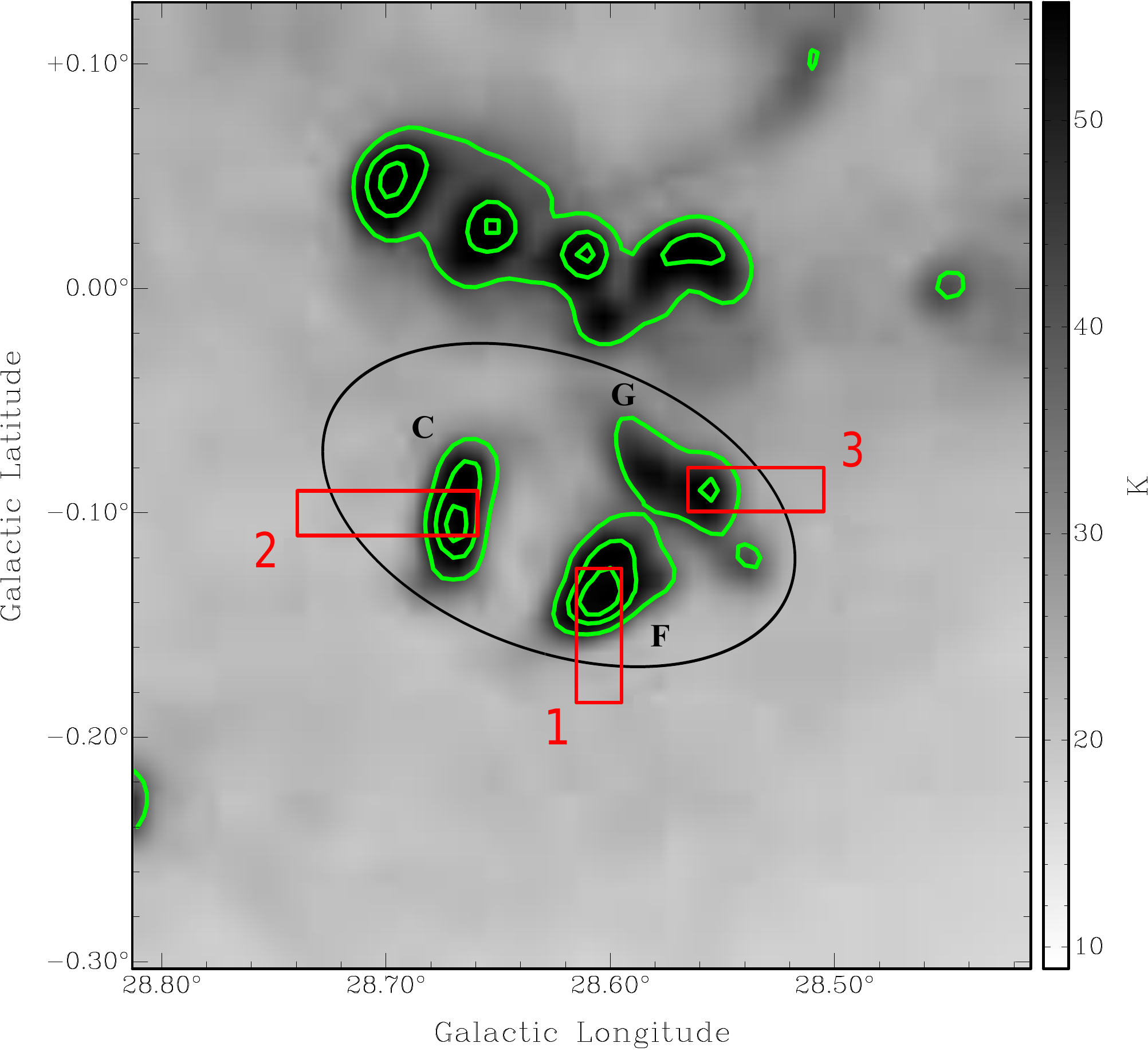}
\caption {SNR G$28.6-0.1$  1420 MHz image with green contours at 40, 60 and 80 K. 
The ellipse is the approximate extent in X-rays of AX J$1843.8-0.352$ given by \citep{Bamba2001}. 
}
\label{fig:g28cont}
\end{figure}
        
\begin{figure}
  \begin{subfigure}{\linewidth}
  \includegraphics[scale=.33]{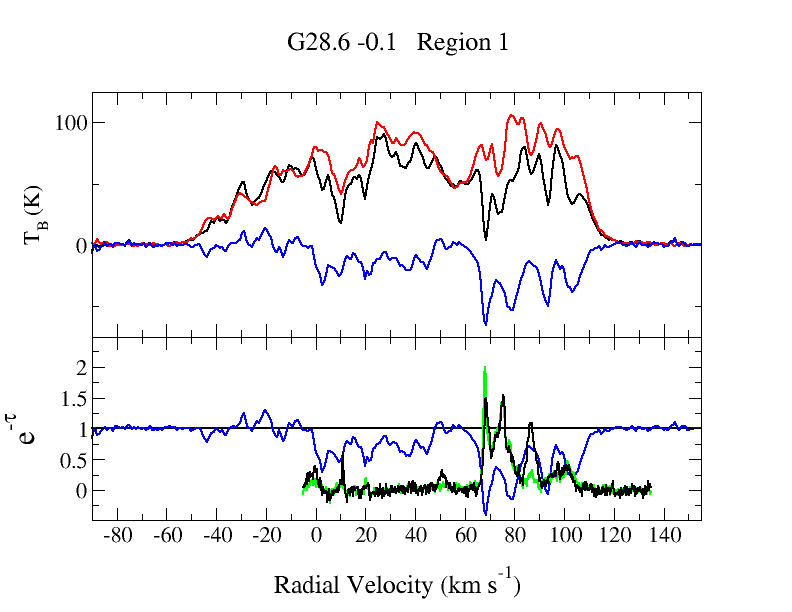}\hfill
  \includegraphics[scale=.33]{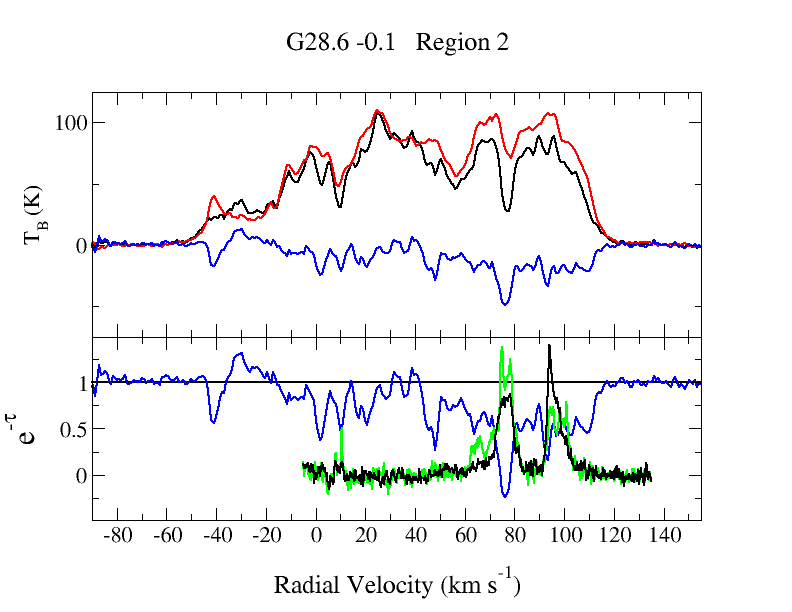}
  \end{subfigure}\par\medskip
  \begin{subfigure}{\linewidth}
  \includegraphics[scale=.33]{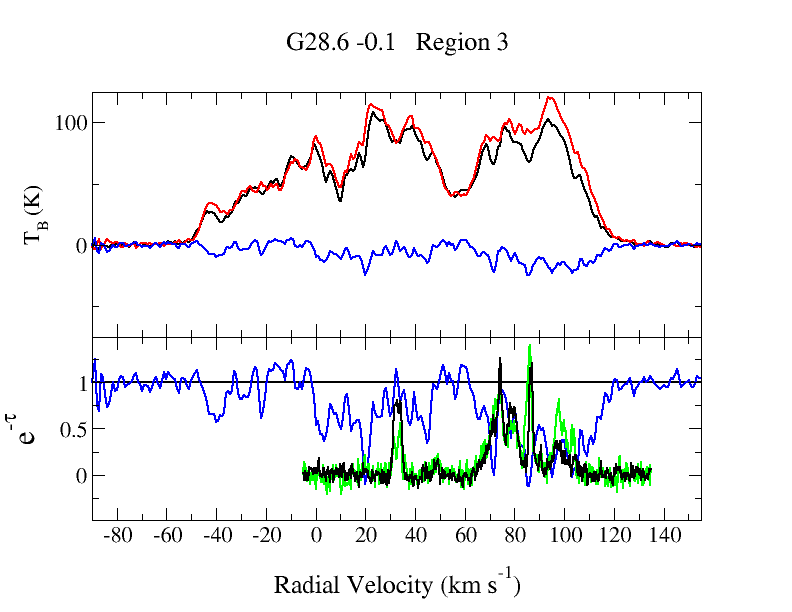}
  \end{subfigure}
\caption{G$28.6-0.1$: Top panel shows the HI emission spectra where the source spectrum (black), background spectrum (red) and difference (blue).
The bottom panel shows the HI absorption spectrum in blue, the $^{13}$CO source spectrum in green and the $^{13}$CO background spectrum in black.}
\label{fig:8}
\end{figure}    

\begin{figure}
\centering
\includegraphics[scale=.38]{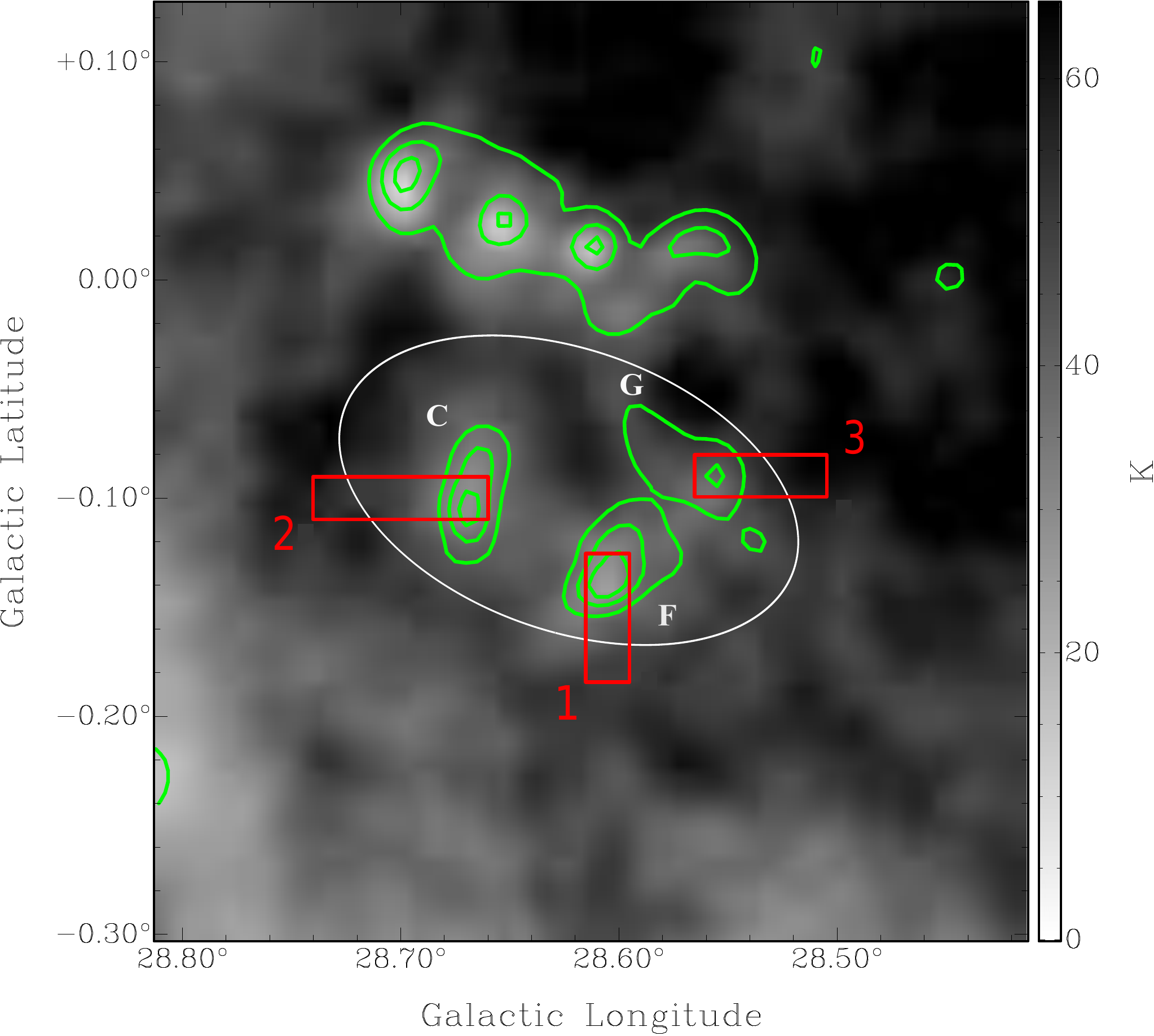}
\includegraphics[scale=.40]{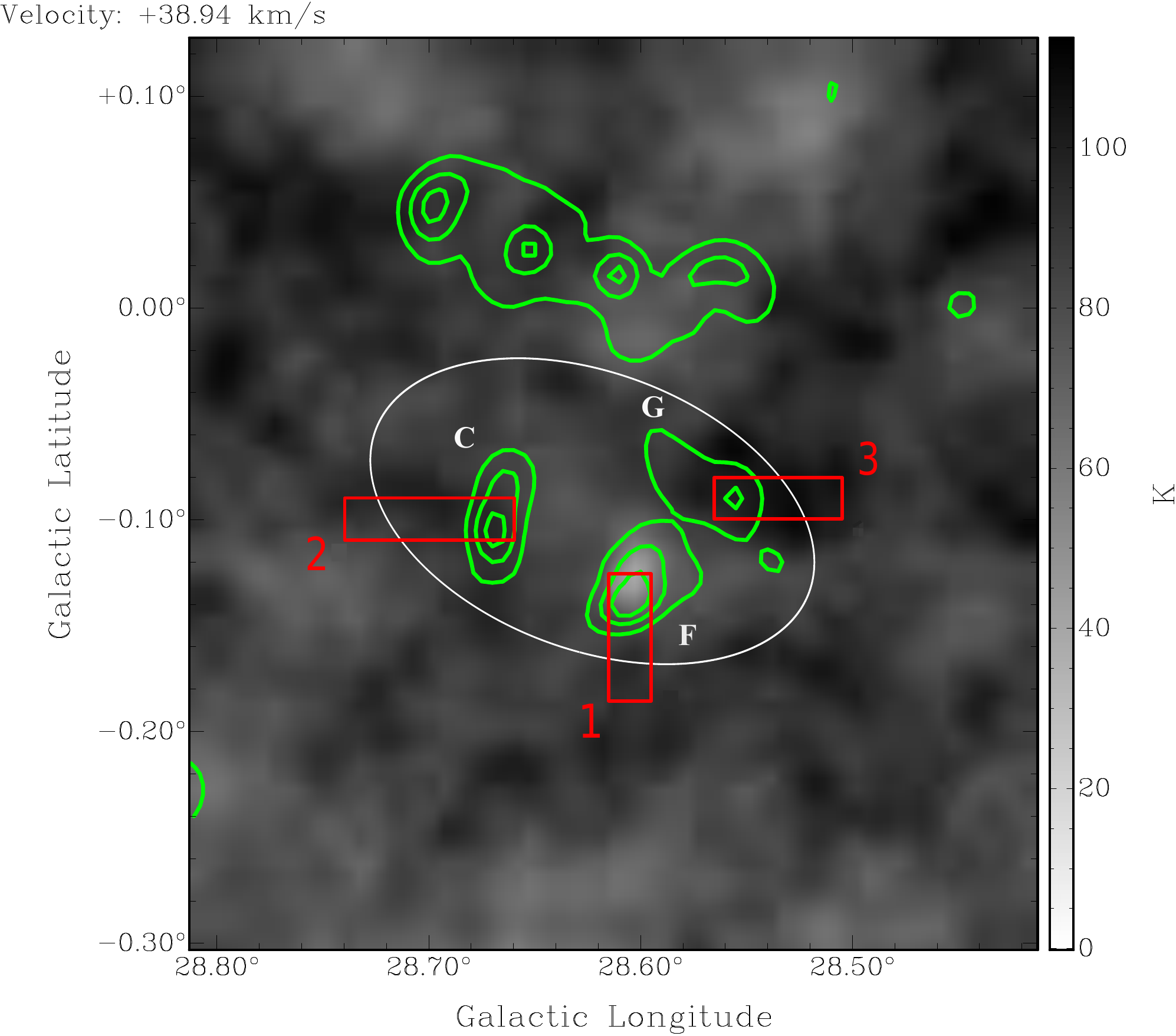}
\caption {G$28.6-0.1$ HI channel maps, top panel: channel maps averaged between 104.89 and 111.49 km s$^{-1}$. Bottom panel: 38.94 km s$^{-1}$ channel map. The green continuum contours at 40, 60 and 80 K.}
\label{fig:9A}
\end{figure}
        
\begin{figure*}
\centering
\includegraphics[width = \textwidth]{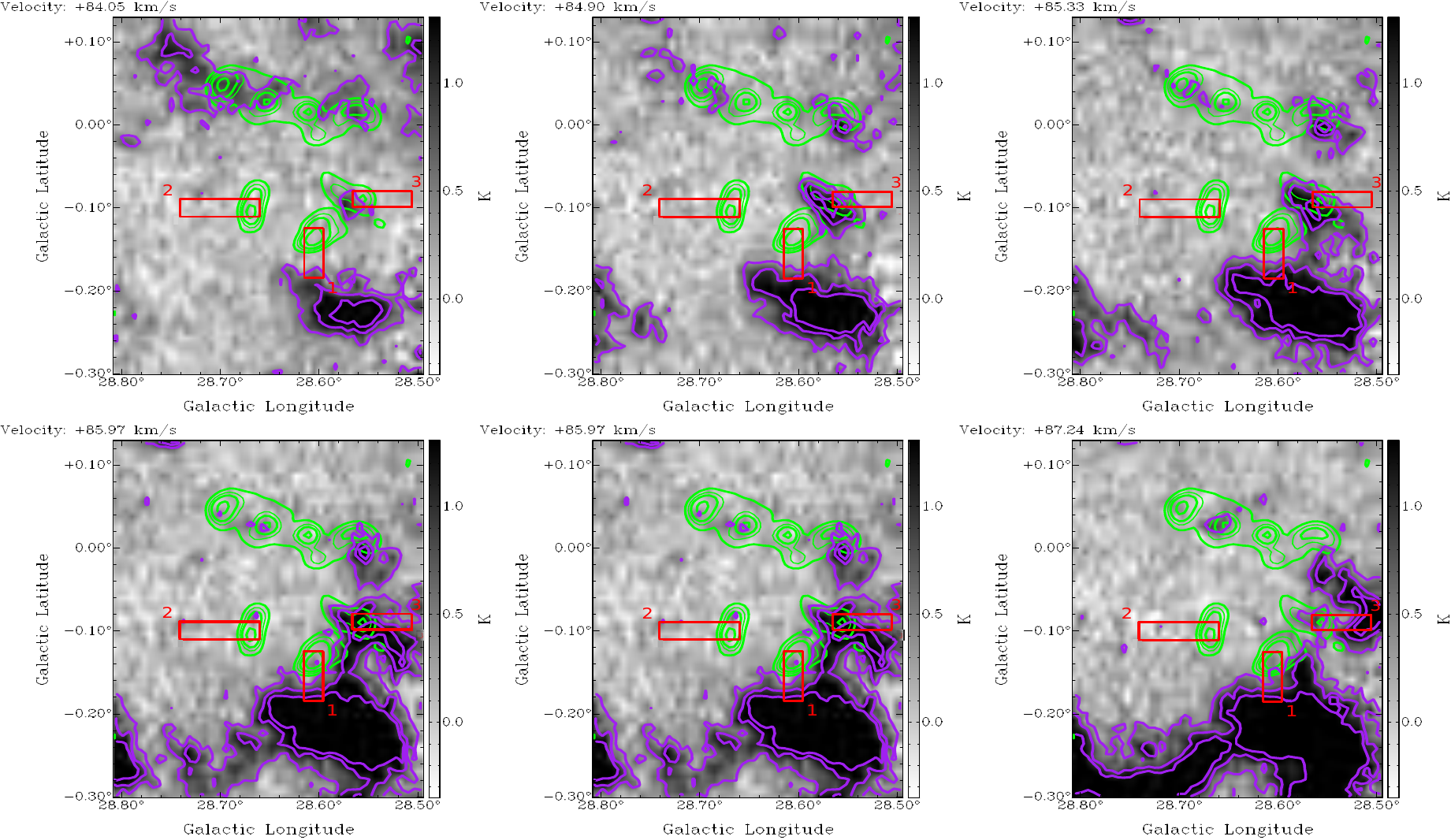}
\caption{G$28.6-0.1$ $^{13}$CO channel maps +84.05 to +87.24 $ \text{km s}^{-1}$. The purple $^{13}$CO contours  at 0.5, 1 and 1.5 K. The green continuum contours at 40, 50, 60 and 80 K.}
\label{fig:9}
\end{figure*}   

\indent G$28.6 -0.1$ is a SNR $13^\prime \times 9^\prime$ in size \citep{2017Green} in the G$28.60 -0.13$ complex. 
\cite{Helfand1989} obtained preliminary classifications for most of the objects in this complex. 
Figure \ref{fig:g28cont} shows the region of the SNR and shows the labels (C, F, G) of 
the radio sources adopted by \cite{Helfand1989}. 
The X-ray morphology for AX J$1843.8-0.352$ has an elliptical shape with a mean diameter of $10^{\prime}$ (\cite{Bamba2001}).
AX J$1843.8-0.352$ is a shell-like SNR which emits predominantly synchrotron X-rays from the shell. 
\cite{Bamba2001} suggested that the non-thermal radio sources  C and F, with spectral indices of $\sim$0.5 and $\sim$0.6 respectively, are associated with the SNR. 
Object G has a flat radio spectrum and is likely an unrelated HII region. 
They estimated the SNR distance to be 6 - 8.5 kpc based on the X-ray column density.

Figure \ref{fig:8} shows the HI and CO spectra which were extracted
for regions 1, 2, and 3 (the red boxes in Figure \ref{fig:g28cont}). 
The HI absorption spectra for the SNR regions 1, 2 and 3 show absorption up to the tangent point velocity of $\sim$110.2 km s$^{-1}$ (Figure \ref{fig:1A}). 
Figure \ref{fig:9A} top panel is an averaged HI image between the velocities 104.89 and 111.49 km s$^{-1}$ and shows absorption in all three sources. 
Thus all three radio components (C, F and G) are located beyond the tangent point distance of 7.3 kpc. Furthermore, the HII region G could be related to the SNR as well.
The channel maps also show that the HI absorption features at negative velocities are false features and correspond to excess HI emission in the background regions. 

Between +34 and +39 km s$^{-1}$ apparent absorption is seen in F but not in C or G. 
From examination of the channel maps it is seen that the feature at $\sim39$ km s$^{-1}$ 
is located slightly, but significantly, above the brightest region of F (Figure \ref{fig:9A} bottom panel). In addition, it changes position with velocity. 
This leads to the conclusion that this absorption is due to:
i) a small dense cool cloud with an offset from the center of
component F and with a small velocity gradient, giving an absorption feature that does not peak at the same position as the continuum and which shifts slightly with velocity;  
or ii) a hole in a large HI cloud that covers the radio sources in the region and which shifts slightly with velocity. 
We favour the former scenario because of the following: 
the dip in HI intensity at the position of F is larger than the other dips in the area (Figure \ref{fig:9A} bottom panel) more 
consistent with absorption than with low HI density at that
velocity; the geometry required is less probable for the case
of a hole than a small absorbing cloud.
For the case of absorption +39 km s$^{-1}$, it is likely
caused by a small cloud at the near distance, so does not appear
in any other component than F.
In either case, the feature seen at +39 km s$^{-1}$ does not affect
the conclusion of the same distance, beyond the tangent point,
for the three components
C, F and G. 

Figure \ref{fig:9} shows the $^{13}$CO channel maps between 84 and 88 km s$^{-1}$. 
These show a molecular cloud which appears to surround the radio source F. 
The morphology of the cloud suggests interaction with the SNR at its lower-right border. 
The central velocity of the molecular cloud of 86 $ \text{km s}^{-1}$  yields a distance of 9.6 kpc to the SNR. \\
\indent We use the absorption spectrum to estimate the HI column density to the SNR, which can be compared to the column density measured in X-rays.
Integrating the HI absorption profile and using an average spin temperature ($T_{}s$) of 135 K \citep{Binney1998}, we find the column density to be $1.6 \times 10^{22}$ cm$^{-2}$.
This value is for HI only and omits any column density in molecular hydrogen.
Alternately, we can estimate the total column density from the distance, assuming a constant density of $n_{H} = 1$ cm$^{-3}$. 
This gives a column density of $3.0 \times 10^{22}$ cm$^{-2}$. 
Both of these are in consistent, within uncertainties, with the column density derived from X-rays by \cite{Bamba2001} of $(2.4-4.0) \times 10^{22}$ cm$^{-2}$.

\section{Discussion and Conclusion} \label{sec:Disconc}

\indent
For the three SNRs studied here, we have found probable molecular cloud associations.
For each SNR, this allowed narrowing the range of possible radial velocities 
to a value consistent with that of the associated molecular cloud. 
The HI absorption spectra and HI channel maps allowed resolution of the Kinematic
Distance Ambiguity. 
The resulting radial velocities, distances and uncertainties in distances for the three SNRs are given in Table \ref{table1}.\\
\indent 
With the distances for the SNRs, we know the sizes and can   
estimate their ages.
The mean radius in the 1420 MHz radio continuum image is taken as the shock wave radius, $R$.
For each  SNR, we apply a basic Sedov model (\cite{1972Cox}) because we don't have enough
data to justify more complex models. 
The shock wave radius in the
Sedov stage is given by $R=12.9$pc$(\epsilon_0/n_0)^{1/5}t_4^{2/5}$, where $\epsilon_0$ is the explosion energy in units 
of $0.75\times10^{51}$erg, $n_0$ is the interstellar medium density in units of cm$^{-3}$, and $t_4$ is the SNR age in units of $10^4$yr. 
We use a mean explosion energy of $5\times10^{50}$ erg, which was found 
for Large Magellanic Cloud SNRs by \cite{2017Leahy}. 
An ISM density of 1 cm$^{-3}$ is used. We note that the radius is sensitive to the
1/5 power of energy/density, and age is sensitive to the 1/2 power of energy/density,
so that the estimates depend weakly on the energy and density assumptions.
The mean angular radius of each SNR, measured from the 1.4 GHz radio continuum image, is 
3.6 arcmin for  G$20.4+0.1$, 
9 arcmin for G$24.7-0.6$ and 5.5 arcmin for G$28.6-0.1$.
Using our distances, we obtain shock radii $R$=7.9 pc for   G$20.4+0.1$, 
$R$=9.9 pc for G$24.7-0.6$ and $R$=15.4 pc for G$28.6-0.1$.
The resulting age estimates from the Sedov model for the three SNRs, respectively, are 3640 yr, 6400 yr and 19000 yr. \\
\indent In this work, we analyse HI spectra, $^{13}$CO spectra, HI channel maps and $^{13}$CO channel maps of three SNRs in the VGPS survey area. 
Previous distance measurements for G$24.7-0.6$ and G$28.6-0.1$ were estimated based on HII region association and X-ray column density respectively.
For each of the SNRs a probable molecular cloud association is found, based on 
morphological match of the SNR boundary with the molecular cloud boundary.
We find distances to the SNRs G$20.4+0.1$, G$24.7-0.6$ and G$28.6-0.1$ as $7.8 \pm0.5$ kpc, $3.8\pm 0.2 $ kpc and $ 9.6 \pm 0.3 $ kpc, respectively.  
From a Sedov model we estimate the ages of the SNRs and find them to be middle aged SNRs ($\sim$4000 to 20,000 yr old).

\begin{table*}
\caption{Distances to supernova remnants}
\centering
\begin{tabular}{ccclccc}
\toprule
\midrule
   & & & & & \\
$\# $ & Source  & Previous     & Reference  & V$_{r} $    & KDAR$^{1}$ & New \\
      &         & Distance(kpc) &             & km s$^{-1}$ &     & Distance (kpc)     \\
  & & & & & \\      
\midrule
\midrule
   & & & & & \\ 
\ref{G20} & G20.4  +0.1  & - & - &  $\sim 120$  &  TP  & $7.8 \pm0.5  $ \\
\ref{G24} & G24.7  -0.6  & $9.3~~$ $^{2}$   & \cite{KooHeiles1991}   & 60.67   &  N  & $3.8\pm 0.2  $  \\     
\ref{G28} & G28.6  -0.1  &  $6 - 8.5~~$ $^{3}$& \cite{Bamba2001} & 86   & F   &  $ 9.6 \pm 0.3 $\\
   & & & & & \\
\midrule 
\bottomrule
\end{tabular}
\begin{minipage}{16cm}
\footnotesize
$\quad$\\
Notes: \\[0pt] 
1. KDAR- Kinematic Distance Ambiguity Resolution, indicating whether the SNR is at the near (N), far (F) or tangent point (TP) distance. \\[0pt]
2. From association with HII region.\\[0pt]
3. From X-ray column density. \\[0pt]

\label{table1}
\end{minipage}   
\end{table*}


\section*{Acknowledgements}

This work was supported in part by a grant from the Natural Sciences and Engineering Research Council of Canada. We thank the referee for insightful comments and suggestions that have improved this work.




\bibliographystyle{mnras}
\bibliography{references} 




\bsp	
\label{lastpage}
\end{document}